%Paper: hep-ph/9211222
%From: luke@yukawa.UCSD.EDU
%Date: Fri, 06 Nov 92 11:04:15 PST

\input harvmac
\def\upsi{{\hbox{``}\Upsilon\hbox{''}}}
\def\subupsi{{\hbox{\sevenrm ``}\Upsilon\hbox{\sevenrm ''}}}
\def\subsubupsi{{\hbox{\fiverm ``}\Upsilon\hbox{\fiverm ''}}}
%
% preprint number
%
\def\UCSD#1#2{\noindent#1\hfill #2%
\bigskip\supereject\global\hsize=\hsbody%
\footline={\hss\tenrm\folio\hss}}% restores pagenumbers
%
% abstract
%
\def\abstract#1{\centerline{\bf Abstract}\nobreak\medskip\nobreak\par #1}
%
%
% titlefont
%
%
\edef\tfontsize{ scaled\magstep3}
 \tfontsize  \tfontsize
 \tfontsize \font\titlei=cmmi10 \tfontsize
\font\titleis=cmmi7 \tfontsize \font\titleiss=cmmi5 \tfontsize
\font\titlesy=cmsy10 \tfontsize \font\titlesys=cmsy7 \tfontsize
\font\titlesyss=cmsy5 \tfontsize  \tfontsize
\skewchar\titlei='177 \skewchar\titleis='177 \skewchar\titleiss='177
\skewchar\titlesy='60 \skewchar\titlesys='60 \skewchar\titlesyss='60

\hbadness=10000

\noblackbox
\vskip 1.in
\centerline{{
\titlefont{Analyticity and the Isgur-Wise Function\footnote{*}{{\tenrm Work
supported in part by the Department of Energy under contracts
DE--AC03--76SF00515 (SLAC), DE--FG03--90ER40546 (UC San Diego) and
DE--AC03--81ER40050 (Caltech).}}}}}
\vskip .5in
\centerline{Adam F.~Falk}
{\it\centerline{Stanford Linear Accelerator Center}
\centerline{Stanford CA 94309}}
\bigskip
\centerline{Michael Luke}
{\it\centerline{Department of Physics 0319}
\centerline{University of California at San Diego}
\centerline{La Jolla CA 92093--0319}}
\bigskip
\centerline{Mark B.~Wise}
{\it\centerline{Department of Physics}
\centerline{California Institute of Technology}
\centerline{Pasadena CA 91125}}

\vskip .3in

\abstract{
We reconsider the recent derivation by de Rafael and Taron of bounds on
the slope of the Isgur-Wise function. We argue that one must be careful
to include cuts starting below the heavy meson pair production
threshold, arising from heavy quark-antiquark bound states, and that if
such cuts are properly accounted for then no constraints may be
derived.
}

\vfill
\UCSD{\vbox{
\hbox{SLAC-PUB-5956}
\hbox{UCSD/PTH 92-35}
\hbox{CALT-68-1830}}}{October 1992}
\eject

It has recently been argued by de Rafael and Taron \ref\derafael{E. de
Rafael and J. Taron, {\sl Phys. Lett.} B282 (1992) 215.} that one may
use analytic properties to derive
constraints on the Isgur-Wise function $\xi(w=v\cdot
v')$ \ref\isgurwise{N. Isgur and M.B. Wise, {\sl Phys. Lett.} B237
(1990) 527.} of the heavy quark effective theory (HQET).
These constraints
have provoked much interest, because they are in conflict with almost
all extractions of $\xi(w)$ from fits to experimental data on
semileptonic $B\to D^*$ decays \ref\neubert{M. Neubert, {\sl Phys.
Lett.} B264 (1991) 455.}, as well as with estimates from QCD sum rules
\ref\sumrules{M. Neubert, {\sl Phys. Rev.} D45 (1992) 2451.} and
potential models \ref\models{N. Isgur and M.B. Wise, {\sl Phys. Rev.}
D43 (1991) 819\semi M. Neubert and V. Rieckert, {\sl Nucl. Phys.} B382
(1992) 97.}.  Hence it is very important to understand whether the
derivation of these bounds is correct.  We have reconsidered this
issue, and we have found that the careful inclusion of heavy
quark-antiquark bound states, lying below the heavy meson pair
production threshold, invalidates the argument. In fact, we argue that
it is not possible to obtain any such constraints at all.

We begin by reviewing carefully the derivation of de Rafael and Taron,
paying particular attention to a number of delicate points which are
somewhat glossed over in their analysis.  We will then present two
counterexamples, namely Isgur-Wise functions which we argue are
reasonable in certain physical limits of the theory, but which violate
the bounds which are derived.  Finally, we will show precisely at which
point the argument fails, and argue that the failure can be understood
both on technical and physical grounds.

For concreteness, we will follow ref.~\derafael\ and take the heavy
quark to be the b quark.  Then the physical form factor $F(q^2)$ which
is of interest is given by the matrix element of the vector current
$V^\mu=\bar{\rm b}\gamma^\mu {\rm b}$ between $B$-meson states:
\eqn\fdef{\langle B(p')|\,V^\mu\,|B(p)\rangle=(p+p')^\mu F(q^2)\,.} For
$q^2\le0$, this is the elastic form factor, corresponding to the
kinematic region described in the HQET by the Isgur-Wise function
$\xi(w)$.  For $q^2\ge4m_B^2$, $F(q^2)$ describes $s$-channel
production of $B\bar B$ pairs.  Conservation of b-number in QCD yields
the normalization condition $F(0)=1$; the question is whether $F(q^2)$,
and hence the Isgur-Wise function, can be further constrained in the
physically interesting elastic region $q^2\le0$.

One begins the argument by considering the two-point function
\eqn\twopoint{\Pi(q^2)\,(q^\mu q^\nu-g^{\mu\nu}q^2)={\rm i}
    \int{\rm d}^4x\,{\rm e}^{{\rm i}q\cdot x}\,\langle 0|\,T\,
    \{V^\mu(x),V^\nu(0)\}\,|0\rangle\,.}
The derivative of $\Pi(q^2)$ satifies an unsubtracted dispersion
relation ($Q^2=-q^2$),
\eqn\disperi{\chi(Q^2)\equiv -{\partial\Pi(Q^2)\over\partial Q^2}
    = {1\over2\pi{\rm i}}\int_C{\Pi(t)\over(t+Q^2)^2}\,,}
where the contour $C$ runs below and then above the physical cut on the
real axis, closed by a circle at $|t|=\infty$.  In the real world, this
cut starts at the mass of the lightest state which couples to the
current $V^\mu$. Ignoring weak and electromagnetic interactions, this
is at $t=4m_\pi^2$, corresponding to the annihilation of the current
into two pions.  It is convenient, however, to suppress annihilation
diagrams, by imagining that $V^\mu$ is actually a flavor-changing
current, between two different, but degenerate, heavy quarks.  Since
these fields are subject to a heavy quark flavor symmetry, none of the
low-energy properties of the theory, in particular no properties of the
Isgur-Wise function, are affected by this change.  But now the cut in
the two-point function starts with a pole at the mass of the lightest
(and stable) $\upsi$ state, $t=m^2_\subupsi$, followed by a series of
cuts associated with the production of additional hadrons.  There is
then a cut starting at $t=4m^2_B$ corresponding to pair production of
$B\bar B$ mesons, and so forth.

There is no contribution from the circle at infinity; then the integral
may be expressed in terms of the discontinuity across the cut, plus an
additional contribution from the $\upsi$ pole,
\eqn\disperii{\chi(Q^2)=\chi_{\rm pole}(Q^2)+
    \int_{(m_\subsubupsi+2m_\pi)^2}^\infty
    {\rm d}t\,{{\rm Im}\,\Pi(t)\over\pi(t+Q^2)^2}\,.}
Because of the optical theorem, the value of ${\rm Im}\,\Pi(t)$ is
proportional to the cross-section for production of on-shell states
with invariant mass $\sqrt{t}$. All physical states which couple to the
current $V^\mu$ contribute to ${\rm Im}\,\Pi(t)$.  Dividing these
states into those lying above and below the continuum $B\bar B$
threshold, we write
\eqn\disperiii{\chi(Q^2)=\chi_{\rm pole}(Q^2)+
    \int_{(m_\subsubupsi+2m_\pi)^2}
    ^{4m_B^2}{\rm d}t\,{{\rm Im}\,\Pi(t)\over\pi(t+Q^2)^2}
    +\int_{4m^2_B}^\infty{\rm d}t\,{{\rm Im}\,\Pi(t)\over
    \pi(t+Q^2)^2}\,.}
The first two terms in eq.~\disperiii\ are positive; if we estimate them
as a sum over perturbative Coulomb bound states, they are proportional
to $\alpha_s^3(m_{\rm b})$ They will not play an important role in the
analysis.  The third term is bounded from below by the contribution of
the $B\bar B$ final states. (Note that there is {\it no\/} relation
between ${\rm Im}\,\Pi(t)$ and $|F(t)|^2$ in the physically
inaccessible region $t<4m^2_B$.  This is why we must make the
decomposition \disperiii.)  The first two terms are positive (and
small) and so, performing an integral over the available phase space,
one then obtains
\eqn\inequality{16\pi^2m_B^2\chi(Q^2)
    \ge{1\over12}\int_1^\infty\,{\rm d}\,y\,{(y-1)^{3/2}\over y^{3/2}
    (y+Q^2/4m_B^2)^2}\,|F(y)|^2\,,}
where $y=t/4m_B^2$.

In the strict limit $m_{\rm b}\to\infty$, the form factor $F(y)$ is
analytic everywhere except along the cut $y\ge1$.  (For finite $m_{\rm
b}$ this is no longer true, and soon we will see that one must treat
this issue with care.) If the integral around the end of the cut can be
neglected, the integral in \inequality\ may be rewritten as a contour
integral, running from $y=\infty$ to $y=1$ along the bottom of the cut,
back to $y=\infty$ along the top, and closing with a circle at
$|y|=\infty$. Then the inequality \inequality, plus the three conditions

(i) $F(y)$ analytic everywhere inside the contour,

(ii) the normalization $F(0)=1$,

(iii) $-Q^2<4m_B^2$ real,

\noindent allow one to derive constraints on $F(y)$, in particular in
the elastic region $y\le0$ \ref\derafael{C. Bourrely, B. Machet and E.
de Rafael, {\sl Nucl. Phys.} B189 (1981) 157.}.  This analysis is
performed most conveniently after the change of variables
$\sqrt{y-1}={\rm i}\,{1+z\over1-z}$, under which the interior of the
contour is mapped onto the open unit disk, and the elastic region
$-\infty<y\le0$ onto the real axis $0\le z<1$.  The constraints take
the form $F_-(z)\le F(z)\le F_+(z)$, where
\eqn\constraints{\eqalign{
    F_\pm(z)=&{1\over(1+z)^2\sqrt{1-z}}\left({1+z+(1-z)
    \sqrt{1+Q^2/4m_B^2}\over1+\sqrt{1+Q^2/4m_B^2}}\right)^2\cr
    &\times \left\{1\pm\sqrt{z^2\over1-z^2}\sqrt{6144\,\pi\, m_B^2
    \,\chi(Q^2)
    \,\left(1+\sqrt{1+Q^2/4m_B^2}\over2\right)^4-1}\right\}\,.\cr}}
If we choose $|Q^2|\gg\Lambda_{\rm QCD}^2$, we may approximate
$\chi(Q^2)$ by its value in perturbation theory,
\eqn\chipert{\chi(Q^2)={3\over4\pi^2}\int_0^1{\rm d}x\,
    {2x^2(1-x)^2\over m_{\rm b}^2+x(1-x)Q^2}+{\cal O}(\alpha_s(Q^2))\,.}
Then expression \constraints\ simplifies to a given function of $z$ and
$Q^2$.

Thus one obtains constraints on $F(z)$ for any value of $Q^2$ such that
the perturbative calculation of $\chi(Q^2)$ is valid.
Since at leading
order, for $y\le0$, the form factor $F(y)$ is equal to the Isgur-Wise
function $\xi(w=v\cdot v'=1-2y)$, eq.~\constraints\ leads directly to a
family of upper and lower limits on the slope $-\rho^2$ of $\xi(w)$ at
$w=1$.  The lower limits on $\rho^2$ (from $F_+(z)$) are not
interesting, as they lie below the kinematic bound $\rho^2\ge{1\over4}$
\ref\bjlimit{J.D. Bjorken, SLAC preprint SLAC--PUB--5278 (1990),
invited talk given at Les Rencontres de Physique de la Vallee d'Aoste,
La Thuile, Italy.}. As for the upper limit, it is strongest if we take
$|Q^2|\ll m_B^2$, which gives the bound quoted ref.~\derafael,
\eqn\drtbound{\rho^2\le1.42\,.}

However, there are reasons to believe that this bound on the slope of
the Isgur-Wise function cannot be correct.  In particular, one may
construct two simple counterexamples.  The first involves a
hypothetical world in which the light quark in the $B$ meson has a mass
$m_\ell$ such that $m_{\rm b}\gg m_\ell\gg\Lambda_{\rm QCD}$.  In this
case the nonrelativistic quark model (NRQM) \ref\isgw{See, for example,
N. Isgur, D. Scora, B. Grinstein and M.B. Wise, {\sl Phys. Rev.} D39
(1989) 799.} is applicable, as the $B$ meson is analogous to the
hydrogen atom, a weakly bound system with coupling strength
$\alpha_s(m_\ell)$.  The ``charge radius'' $\xi'(1)$ may be computed
using nonrelativistic wavefunctions for the bound states, and one
obtains a value which is proportional to $-1/\alpha_s^2(m_\ell)$, in
violation of the proposed bound.  If such a ``$B$-meson'' were really
like a hydrogen atom, in particular if it were unconfined, the
derivation would have failed because of an anomalous threshold in
$F(q^2)$ at the point $q^2\approx\alpha_s^2m_{\rm b}^2$. Yet the NRQM
should apply arbitrarily well for the confinement scale arbitrarily in
the infrared; for such ``weakly confined'' QCD \ref\howard{See H.
Georgi, {\sl Weak Interactions and Modern Particle Theory}, Benjamin
Cummings Publishing Co., 1984.}, $F(q^2)$ must be analytic all the way
up to the point $q^2=m^2_\subupsi$. However in this case there is still
a ``would-be anomalous threshold'' near $q^2=0$ controlling the steep
behaviour of $\xi(w)$ at zero recoil. In potential models, such
would-be anomalous thresholds are typically associated with the
presence of many poles in $F(q^2)$, with residues large in magnitude
and oscillating in sign, along but near the end of the physical cut
\ref\jaffe{R.L. Jaffe and P.F. Mende, {\sl Nucl. Phys.} B369 (1991)
189.}.  This is a hint that it may be necessary to consider more
carefully the behaviour of $F(q^2)$ near the physical threshold.

For our second counterexample, we recall that the techniques of HQET
may be used to extract the leading logarithmic dependence on the heavy
quark mass $m_{\rm b}$ of the physical form factor $F(y)$. Neglecting
terms of order $1/m_{\rm b}$, we find \ref\fggw{A.F. Falk, H. Georgi,
B. Grinstein and M.B. Wise, {\sl Nucl. Phys.} B343 (1990) 1.}
\eqn\counterii{F(y)=\left[\alpha_s(m_{\rm b})\right]^{a_L(w)}\,
    \xi_{\rm ren}(w)\,,}
where
\eqn\aldef{a_L(w)={8\over33-2N_f}\left[{w\,\ln\left(w+\sqrt{w^2-1}\right)
    \over\sqrt{w^2-1}}-1\right]\,,}
$\xi_{\rm ren}(w)$ is a universal $m_{\rm b}$-independent and
$\mu$-independent function, and $N_f$ is the number of light flavours.
Now for $m_{\rm b}$ arbitrarily large,
$\alpha_s(m_{\rm b})$ is arbitrarily small, and the magnitude of the
slope of $F(y)$ at $y=0$ may be made arbitrarily large (note that as
$m_{\rm b}$ grows, the inequality \inequality\ is better and
better satisfied).  Of course, such behaviour is inconsistent with {\it
any\/} bound on the slope of the form factor which one might hope to
derive.

So where does the argument which leads to the constraints \constraints\
fail?  The problem is that it is not possible to convert the integral
from $y=1$ to $y=\infty$ of $|F(y)|^2$ (times the weighting function)
into a contour integral, {\it because it is not possible to integrate
around the end of the physical cut at $y=1$.}  This is simple to see
when $F(y)$, taken from eq.~\counterii, is expanded about $y=1$,
corresponding to $w=-1$.  Writing $w=-1+\varepsilon$ and taking
$\varepsilon\ll1$, we find
\eqn\essential{F(y)\propto\left[\alpha_s(m_{\rm b})\right]^{a_L(w)}
    \sim {\rm e}^{K/\sqrt{\varepsilon}}\,,}
for some positive constant $K$.  We see that $F(y)$ has an {\it essential
singularity\/} at the point $y=1$.  The integral around the end of the
cut, taken on a small circle of radius $\delta$, has no well-defined
limit as $\delta\to0$; hence it is not possible to find a closed
contour on which the (weighted) integral of $|F(y)|^2$ is bounded, and
inside of which $F(y)$ is analytic.  Therefore condition (i) cannot be
met, and no constraints such as eq.~\constraints\ may be derived.

To understand what has happened, we need to look more carefully at the
analytic structure of $F(y)$.  The integral in the inequality
\inequality\ starts at $y=1$, running to $y=\infty$ along the top of
the cut given by continuum production of $B\bar B$ pairs.  However,
there are additional singularities in $F(y)$, corresponding to diagrams
in which the current $V^\mu$ converts to a heavy quark-antiquark bound
state, which then couples directly to the $B$-meson.  These are
singularities below $B\bar B$ threshold, beginning at the mass of the
$\upsi$ state, which for very large $m_{\rm b}$ is located at
$y=1-4\alpha_s^2(m_{\rm b})/9$. To complete the contour integral in such
a way that $F(y)$ is analytic everywhere inside, one must close it to
the left of $y=1-4\alpha_s^2(m_{\rm b})/9$.  These additional
contributions to the
weighted integral of $|F(y)|^2$ need not be negligible, and they are
not known.  Hence once again we see that we cannot take the crucial
step from a bounded integral over $1\le y<\infty$, to a bounded
integral over a closed contour.

In fact, in both of our counterexamples we should expect singular, or
at least dramatic, behaviour associated with the $\upsi$ region below
$B\bar B$ threshold. In weakly confined QCD, the presence of the
would-be anomalous threshold near $y=0$ is typically associated with
rapid variations in $F(y)$ \jaffe, which it is natural to associate
with the $\upsi$ region.  In the HQET, the singular leading logarithmic
behaviour \counterii\ of $F(y)$ at $y=0$ arises from the infrared
divergence associated with multiple gluon exchange between heavy quarks
with nearly the same velocity.  The quark and antiquark  would like to
form a bound state, but cannot in the $m_{\rm b}\to\infty$ limit with
their velocities absolutely fixed.  One may be misled if one fails
to account carefully for the contributions of these bound states {\it
below\/} the threshold for heavy meson pair production.

In conclusion, we wish to stress not simply that the analysis of
ref.~\derafael\ is flawed, but that it is not likely to be possible to
amend this argument so that weaker, but still rigorous, bounds could be
derived.  Of course one could attempt to model the contributions to
$F(q^2)$ of the heavy quark-antiquark bound states, but such an
approach would undermine the rigor of the derivation.  We have
presented two counterexamples in which the magnitude of the slope
$\xi'(1)$ may be made arbitrarily large, and in which any universal
bounds which one might hope to derive would be violated.  We have argued
that
these are sufficient to preclude the derivation of any such universal
constraints on the Isgur-Wise function.

Similar work has also been done by B.~Grinstein and P.~Mende
\ref\grmende{B.~Grinstein and P.~Mende, SSCL-PP-167/BROWN HET-882.}
and by C.~E.~Carlson, N.~Isgur,
T.~Mannel, J.~Milanu and W.~Roberts.  We are grateful to them for
communicating their results
to us prior to publication.  We also thank D.~Kaplan, M.~Neubert,
M.~Peskin and J.~Taron for useful conversations.  A.~F.~would like to
thank the Department of Physics at UC San Diego, where portions of this work
were performed, for their hospitality.

\listrefs

\bye